\documentclass[prl,twocolumn,superscriptaddress,showpacs,preprintnumbers,amsmath,amssymb]{revtex4}
\usepackage{amssymb}

\usepackage{dcolumn}%Align table columns on decimal point
\usepackage{bm}% bold math
\usepackage{epsfig}

\newcommand{\ignore}[1]{}

\begin{document}

\begin{titlepage}

\title{Interplay between Quantum Size Effect and Strain Effect on Growth of Nanoscale Metal Thin Film}

\author{Miao Liu,$^1$ Yong Han,$^1$$^,$\footnote[2]{Current address: Ames Laboratory, 307D Wilhelm Hall, Ames, IA50011} Lin Tang,$^2$ Jin-Feng Jia,$^{2}$ Qi-Kun Xue,$^{2}$ Feng Liu,$^1$$^,$\footnote[1]{Email: fliu@eng.utah.edu}}

\address{$^1$Department of Materials Science and Engineering, University of Utah, Salt Lake City, Utah 84112, USA}
\address{$^2$Department of Physics, Tsinghua University, Beijing 100084, China}

\date{\today}
\begin{abstract}
We develop a theoretical framework to investigate the interplay between quantum size effect (QSE) and strain effect on the stability of metal nanofilms. The QSE and strain effect are shown to be coupled through the concept of \textquotedblleft quantum electronic stress. First-principles calculations reveal large quantum oscillations in the surface stress of metal nanofilms as a function of film thickness. This adds extrinsically additional strain-coupled quantum oscillations to surface energy of \textit{strained} metal nanofilms. Our theory enables a quantitative estimation of the amount of strain in experimental samples, and suggests strain be an important factor contributing to the discrepancies between the existing theories and experiments.

\end{abstract}


\maketitle
\end{titlepage}

When the thickness of a metal film is reduced to the range of electron Fermi wavelength, quantum confinement becomes prominent to form discrete quantum well states, giving rise to various manifestations of quantum size effect (QSE) \cite{F.K.Schulte}. In particular, the QSE has been shown to be a dominant factor in the growth of metal nanofilms on semiconductor substrates \cite{F.K.Schulte,Z.Zhang,V.Yeh,C.M.Wei,M.H.Upton} in the so-called electronic growth regime \cite{Z.Zhang}. On the other hand, the strain effect is ubiquitous in heteroepitaxial growth of semiconductor and metal thin films \cite{F.Liu M.G.Lagally,F.Liu}. A few recent studies \cite{Z.Suo,Jiang,Kuntova,Unal1,Unal2} have considered both effects on metal thin film growth. One thermodynamic theory \cite{Z.Suo} studied both effects on film stability, and two kinetic models \cite{Kuntova,Unal1} assumed growth parameters to be dependent of island height and radius due to the QSE and strain effects. However, majority studies have focused on one effect while neglecting the other, and those few studies which considered both effect have been generally limited to treat them as two independent additive effects. This is mostly because fundamentally no theory is available to assess how the QSE may change the stress state of the film, and conversely how strain may alter the QSE. Therefore, it is very important to establish a theoretical framework that underlies the QSE on surface stress that in turn unerlies the interplay between the QSE and strain effect.

The Pb(111) film grown on Si(111) substrate has been extensively studied as a model system for QSE \cite{V.Yeh,C.M.Wei,M.H.Upton,P.Czoschke,Y.F.Zhang1}. The almost perfect matching between the Pb Fermi wavelength and its interlayer spacing in the (111) direction gives rise to two striking QSE features in Pb film: the odd-even oscillations and beating patterns exhibited in many properties such as surface energy and stability. These two main features have been agreed upon by all theoretical and experimental studies \cite{V.Yeh,C.M.Wei,M.H.Upton,P.Czoschke,Y.F.Zhang1}. However, there remain some outstanding discrepancies. Oscillation patterns may vary slightly from one experimental sample to another \cite{P.Czoschke,Y.F.Zhang1,M.M.Ozer,Y.F.Zhang}. First-principles calculations \cite{C.M.Wei} predicted that the odd-even oscillations in surface energy essentially die out at a thickness of $\sim$20 monolayers (MLs), while experiments, in contrast, have seen the large oscillations sustain even beyond 30 MLs \cite{P.Czoschke, Y.F.Zhang1}. One origin of the discrepancies was attributed to Pb/Si interface that causes a phase shift in the oscillation patterns\cite{V.Yeh}, but the strain effect has been mostly overlooked so far.

Because of the large lattice mismatch, the Pb (111)film tends to grow on Si(111) substrate by adopting a 10-to-9 epi relation to minimize interfacial misfit strain \cite{Hupalo,H.H.Weitering}. Even so, Pb film can still experience up to $\pm$3$\%$ strain depending on the film orientation relative to Si surface \cite{H.H.Weitering}. The measurement of interlayer spacing by X-ray diffraction \cite{P.Czoschke} suggested that the strain in Pb film be small based on bulk Poisson ratio, but the actual amount of in-plane strain remains uncertain, because the ultrathin film may not follow the bulk Poisson ratio, especially in the presence of QSE that modifies the interlayer spacing. Overall, the strain effect has not been studied adequately in relation with the QSE, because of the lack of theory underlying their relationship and because the direct measurement of strain in the film is very difficult.

In this letter, we develop a general theory underlying the fundamental relationship between the QSE and strain effect in the formulation of surface energy through the concept of \textquotedblleft quantum electronic stress\textquotedblright \cite{quantum stress}, i.e. the additional surface stress oscillations induced by the QSE. It allows us to theoretically study the interplay of these two effects on the stability of metal nanofilms by treating both effects on the same footing. Using first-principles calculations, we reveal large quantum oscillations in the surface stress of Pb(111) films as a function of thickness, which adds extrinsically additional strain-mediated quantum oscillations to surface energies of the strained Pb films. Our theory enables a quantitative estimation of the amount of strain in different experimental samples from the measured stability patterns.

We first briefly introduce the concept of quantum electronic stress that gives rise to the quantum oscillations of surface stress. Following density functional theory (DFT), the total energy functional of a solid is written as
\begin{equation}
\label{Eq_1}
 E[n(\vec{r}),\{\vec{R}_m\}]=E_e[n(\vec{r})]+E_{ext}[n(\vec{r}),\{\vec{R}_m\}]+E_{I}[\{\vec{R}_m\}]
\end{equation}

$E_e[n(\vec{r})]$ is the electronic energy functional of charge density $n(\vec{r})$, including kinetic and electron-electron interaction energy, $E_{ext}[n(\vec{r}),\{\vec{R}_m\}]$  is the
ion-electron interaction energy, $E_{I}[\{\vec{R}_m\}]$ is the ion-ion interaction energy and $\{\vec{R}_m\}$  are atomic coordinates. Consider a variation of electron density from the
ground-state $n^0$ as $n^*=n^0+\delta n$ in the absence of strain (i.e., without any lattice deformation), a general expression for lattice stress induced by such pure electronic perturbation/excitation can be derived as \cite{quantum stress}

\begin{equation}
 \sigma^{QE}_{ij}=\frac{1}{V} \int_V \frac{\partial \mu}{\partial \varepsilon_{ij}}\delta n(\vec{r})d\vec{r},
\end{equation}

which is called quantum electronic stress. $\mu$  is electron chemical potential and $\partial \mu/\partial \varepsilon_{ij}$ is electron deformation potential. In a nanofilm of thickness $d$, QSE induces variation of charge density and deformation potential along the surface normal z-direction. Then a special form of quantum electronic "surface" stress can be expressed as

\begin{equation}
 \sigma^{QE}_{ij}=\frac{1}{d} \int \frac{\partial \mu}{\partial \varepsilon_{ij}}\delta n(z)dz.
\end{equation}

We have performed DFT calculations to directly reveal quantum surface stress oscillations in Pb(111) nanofilms. Our calculations are done using VASP code \cite{Kresse} based on density functional theory in plane-wave formalism. For all the freestanding Pb films and Pb film on Si substrate from 1$\sim$11 MLs,  ultrasoft pseudopotential \cite{Vanderbilt} and generalized gradient approximation are used with the Pb 5d orbitals included as valence states. For thicker Pb film (12MLs and thicker) on Si substrate, PBE potential \cite{Perdew} and generalized gradient approximation without 5d orbitals are used to save time. All calculations use a plane-wave cutoff of 240eV to obtain good convergence for stresses which typically converge slower than total energy. The Pb film is modeled by a supercell slab with the strain-free film set at the theoretical lattice constant of 5.04\AA. The Si substrate was modeled using 6 layers of Si with the bottom two layers fixed at bulk positions and the bottom layer passivated with H. The slabs are separated by a vacuum thickness of  >20\AA\ in z-direction, sampled by a 20x20x1 mesh in k-space.

\begin{figure}[tbp]
\includegraphics[width=8.5cm]{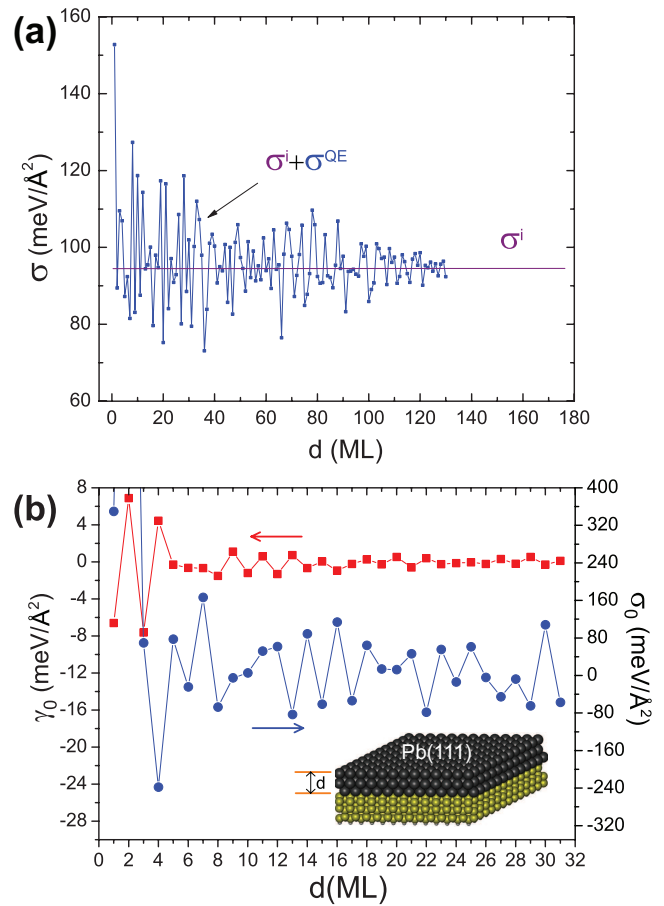}
\caption{Surface energy (squares) and surface stress (dots) of freestanding and Si-supported unstrained Pb(111) film obtained from DFT calcualtions. The insets show schematics of film.}
\end{figure}

Figure 1(a) shows the calculated surface stress ($\sigma_0$), as a function of film thickness ($d$) up to 130 MLs, of the freestanding strain-free Pb(111) film. It is well-known that surface energy displays an oscillatory dependence on $d$ \cite{C.M.Wei}.  What's new is that surface stress $\sigma$ displays also a strong oscillatory dependence on $d$, indicating a strong QSE on surface stress, as suggested by Eq. (3) above. The surface stress exhibits a similar oscillation pattern with film thickness as surface energy \cite{C.M.Wei}, characterized by an odd-even oscillation superimposed by a beating pattern with a period of $\sim$9 layers, but the oscillations of surface stress are out of phase with those of surface energy and the beating patterns are phase shifted too (not shown).

where $E^C$ is the classical surface energy (bond breaking energy) of a macroscopic thick film independent of film thickness, and $E^Q$ is the quantum surface energy due to the QSE, which is function of film thickness $d$. By definition, surface stress (a rank-2 tensor) can be expressed as

In general, we may also express the surface stress as

\begin{equation}
\sigma=\sigma^M+\sigma^{QE}(d) ,
\end{equation}

where $\sigma^M$ is the mechanical surface stress of a macroscopic thick film which we are familiar with, and $\sigma^{QE}$ is the new oscillating quantum surface stress as a function of film thickness $d$. The thickness dependence of the quantum surface stress can be related to the thickness dependence of charge density and electron deformation potential induced by QSE as shown in Eq. (3). As the film thickness increases, however, $\sigma^{QE}$ will eventually diminish and $\sigma$ will converge to $\sigma^M$ as indicated in Fig. 1(a) although we could't calculate thicker film beyond the 130 ML to show full convergence.

The introduction of the quantum surface stress provides a direct link between the QSE and strain effect on the surface energy and hence stability of thin films in the quantum regime. In particular, under a given strain $\varepsilon$, the surface energy will have the following thickness dependence within linear elasticity

\begin{equation}
E(\varepsilon)=E_0(d)+A[\sigma^M+\sigma^{QE}(d)]\cdot\varepsilon,
\end{equation}

Where the first term is the surface energy of a unstrained film (denoted by subscript \textquotedblleft0\textquotedblright) which has a thickness ($d$) dependence  (quantum oscillations) due to the QSE alone, while the second term is the strain induced surface energy which adds extrinsically additional strain-coupled quantum oscillations to surface energy through oscillating quantum surface stress. Equation (5) enables a quantitative assessment of the interplay between the QSE and strain effect on the stability of metal nanofilms.

In experiments, Pb films are grown on semiconductor substrates, such as Si and Ge. Hence, in order to compare with experiments, we must also include the substrate and interfacial effects. Figure 1(b) shows the calculated $\gamma_0$ and $\sigma_0$ as a function of \textit{d} ranging from 1 to 31 MLs of the strainfree Pb(111) film on a Si substrate (To do so, the Si substrate is strained to match the Pb lattice \cite{M.H.Upton}). There are some interesting similarities and differences between the freestanding and substrate-supported Pb (111) film. In both cases, $\gamma_0$ and $\sigma_0$ show an odd-even oscillation modulated by a 9-layer beating pattern; $\sigma_0$ displays a larger oscillation magnitude than $\gamma_0$. On the other hand, for the substrate-supported film, $\gamma_0$ and $\sigma_0$ contain not only the surface terms but also the interface terms, which makes their oscillation amplitude almost twice as large as that of the freestanding film. In addition, the presence of Si substrate causes a phase shift in $\gamma_0$ and $\sigma_0$ by $\sim$1 ML; the first node of the beating pattern appears at the 7ML and 6ML for the freestanding and Si-supported Pb film, respectively.

\begin{figure}[tbp]
\includegraphics[width=8.5cm]{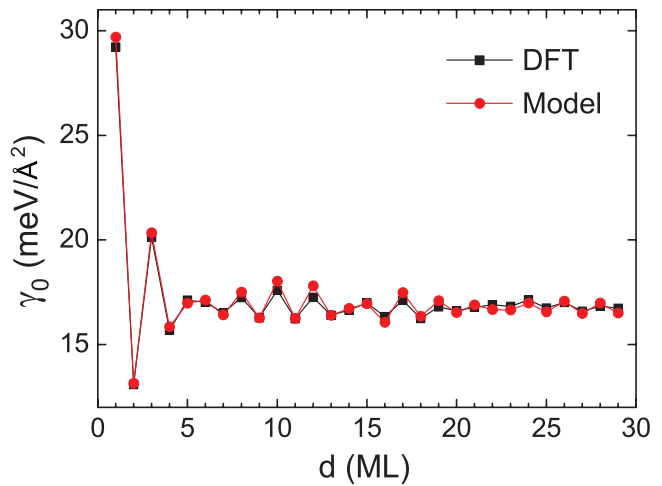}
\caption{Comparison of surface energy between model prediction and direct DFT calculation for a Pb film under 1\% strain, showing excellent agreement. }
\end{figure}

To verify our theoretical framework, we calculated the surface energies of the 1\% strained film as a function of thickness in comparison with the model predictions, as shown in Fig 2. We see that the model predictions agree very well with the direct first-principles results, validating our theory. Thus, using the first-principles calculated the surface energies and surface stresses of the "unstrained" film, we can apply our model to predict the surface energy ($\gamma$) of the strained film with or without substrate support.

Figure 3(a)and(b) shows the predicted surface energy of the freestanding and Si-supported Pb(111) films strained from -3\% to 3\%, respectively. Strain modifies the surface energy in two important ways. First, strain enhances the QSE by increasing the odd-even oscillation magnitude in $\gamma$. This enhancement extends the QSE induced surface energy oscillations to much thicker films (the oscillation persisting beyond 30 ML with $\sim$3\% strain). So, strain provides one possible reason for the experimentally observed stability oscillations existing in much thicker films (>30 ML) \cite{P.Czoschke} than the previous theoretical predictions ($\sim$20 ML) \cite{C.M.Wei}. Second, because the quantum oscillations in surface stress and surface energy are phase shifted, large enough strain will change the oscillation pattern (both the odd-even and beating pattern) of surface energy. This means that strain will alter the relative film stability of different thicknesses. For example, for the strain-free freestanding film, the 14ML film is stable and the 15ML is unstable; however, under 3\% strain, the 14ML becomes unstable and the 15ML becomes stable, as shown in Fig 3(a).

Experimentally, the observed stability patterns of Pb(111) films grown on Si(111) from different groups are in generally good agreement but with some subtle differences around the nodal points of thicknesses in the beating pattern \cite{P.Czoschke,Y.F.Zhang1,M.M.Ozer,Y.F.Zhang}. The reason for such discrepancy remains unresolved, although some general argument has been made by attributing the discrepancy to nonspherical Fermi surface \cite{A.Ayuela} and substrate effect \cite{Z.Suo,V.Yeh}. Here, we argue that the discrepancy is partly caused by the different amount of strain in different experimental samples. Below, we apply our model to extract the amount of strain in some experimental samples by matching the predicted stability patterns to the experiments.

\begin{figure}[tbp]
\includegraphics[width=8.5cm]{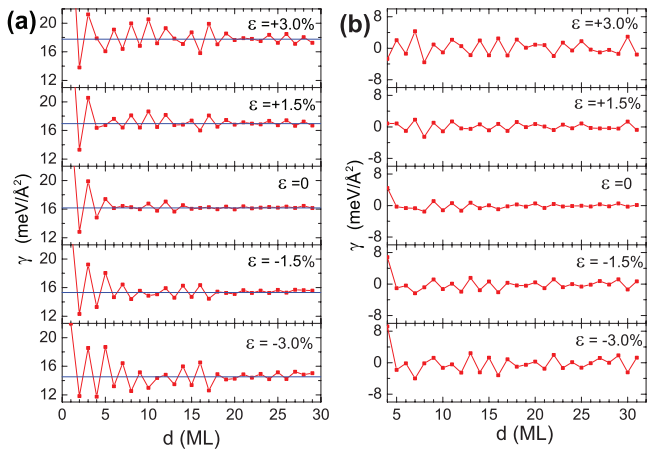}
\caption{Model predicted surface energies of Pb(111) films under strain from -3\% to 3\%. (a)freestanding film and (b)Si-supported film.}
\end{figure}

\begin{figure}[tbp]
\includegraphics[width=8.5cm]{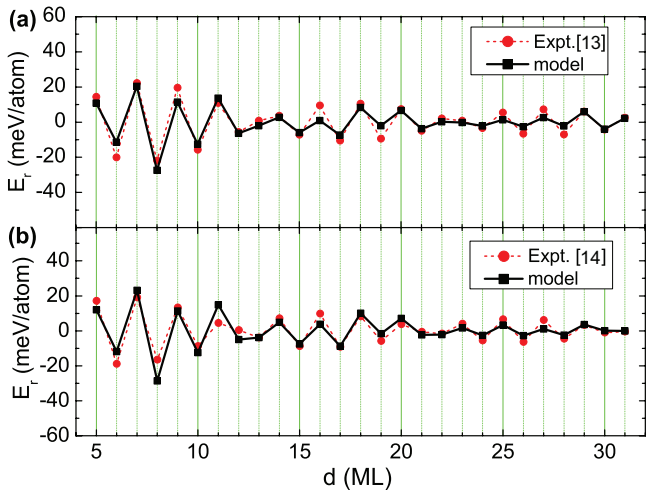}
\caption{Comparison of relative surface energies of Pb(111) film on Si substrate between the experiment (dots) \cite{P.Czoschke, Y.F.Zhang1} and model prediction with the fitted amount of strain (square).}
\end{figure}

Without strain, the calculated stability pattern from either freestanding or Si supported film agrees poorly with the experiment by Czoschke \cite{P.Czoschke} and Zhang\cite{Y.F.Zhang1}, as seen by comparing Fig. 1 with Fig. 4. In particular, both experimental results show large odd-even oscillations from 5 to 8 ML (Fig. 4), while the theory shows little oscillation in this region (Fig. 1) which is in the vicinity of a nodal point of the beating pattern. To resolve this discrepancy, we apply the above theoretical framework [Eq.(5)] to predict the stability pattern of "strained" Pb films on the Si substrate, using the calculated surface/interface energies and stresses of the unstrained film on the Si substrate. In fitting the experimental data, we assume a nonuniform strain distribution in the film that decreases linearly with the increasing film thickness \cite{supplementary material,M.H.Huang}, and then treat the strain and its decay rate as fitting parameters. We obtained very good fitting results by using a linear strain profile of $1.76\%-(d-5)\times 0.068\%$ for Czoschke's sample \cite{P.Czoschke} and $1.80\%-(d-5)\times 0.061\%$ for Zhang's sample \cite{Y.F.Zhang1}, respectively, as shown in Fig.4. Most noticeably, our model correctly predicted the large odd-even oscillations in the range of 5-8 ML as seen in the experiments. This is because there is a large oscillation in the surface stress in this range (see Fig.1), which induces additional oscillations in surface energy when strain is applied. The fitted strain are only slightly different in the two samples by $\sim 0.1\%$, in accordance with the overall agreement between the two experimental patterns. Surprisingly, this small difference is enough to account for the subtle differences in the two experimental patterns in the thickness range of 12-14 ML, 21-23 ML and 30-31 ML, all in the vicinity of nodal points. Overall, the strain is small, less than 2\% initially, and decays with the increasing film thickness to less than 1\% beyond 10 ML and diminishes around 30 ML. The average strain in a 30 ML film is $\sim$ 0.9\%, within the range of general estimation \cite{H.H.Weitering}.

Recently, Miller, et al. has shown a fundamental phase relationship between the oscillations of surface energy and of work function that their beating patterns are always offset by 1/4 of a period \cite{T.Miller}. We have shown that the strain can not only change the odd-even oscillations but also shift the phase of beating patterns of surface energy \cite{Hong}. Applying the Miller's phase relation to the Si-supported Pb(111) film by assuming that the interface shifts the work function and surface energy phase together \cite{Chiang}, we can fit the phase of surface energy beating pattern to match (by an offset of 1/4 of a period) the experimental phase of work function pattern, such as the one measured by Qi et al.\cite{Y.Qi}, using strain as a fitting parameter. We obtained the best fit with an average 0.75\% strain for this particular film, as shown in Fig. 5.

\begin{figure}[tbp]
\includegraphics[width=8.5cm]{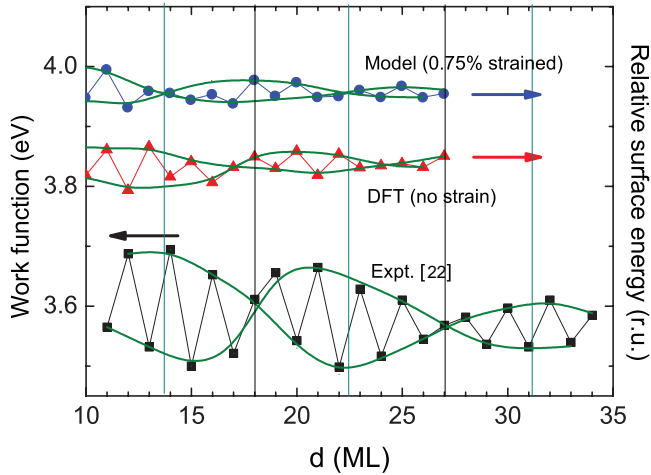}
\caption{Comparison of experimental work function pattern \cite{Y.Qi} with DFT calculated surface energy pattern without strain and
with model predicted surface energy pattern with 0.75\% strain. Note the 1/4 of a period of phase shift between the experimental data and model prediction.}
\end{figure}

In conclusion, we have developed a theoretical framework to investigate the interplay between QSE and strain effect on the thermodynamic stability of metal nanofilms, through the introduction of a new concept of \textquotedblleft quantum stress\textquotedblright \cite{quantum stress}. In the present case, the quantum stress represents the additional surface stress induced by QSE. Broadly, our theoretical framework can be extended to investigate the interplay between QSE and strain effect on kinetic growth properties, such as surface diffusion and step-edge barrier, where quantum "diffusional" stress \cite{Su} induced by the QSE can be derived from first-principles to play the role of quantum surface stress here. Thus, our theory will be applicable to both thermodynamic and kinetic properties of nanoscale thin films when QSE and strain effects are prominent.

M. Liu and F. Liu acknowledge support by NSF (Grant No. DMR0909212) and thank Q.K. Xue for hosting their visit of Tsinghua University as part of NSF-MWN program. Y. Han thanks support by DOE-BES (Grant No. DE-FG02-04ER46148). We thank CHPC at University of Utah and DOE-NERSC for providing the computing resources.

\end{document}